\documentstyle[12pt]{article}
\def\bq{\begin{equation}}
\def\eq{\end{equation}}
\def\bqn{\begin{eqnarray}}
\def\eqn{\end{eqnarray}}

\def\m{\mu}

\def\r{\sqrt{2}}
\def\f{\phi}
\def\t{\theta}

\begin{document}
\thispagestyle{empty}
\begin{flushright}
{\sl CINVESTAV-FIS-02/95}
\end{flushright}
\vspace{1cm}
\begin{center}{\small\bf CONSTRAINTS ON $Z_1$ - $Z_2$ MIXING FROM THE DECAY
$Z_1\rightarrow e^-e^+$ IN THE LEFT-RIGHT SYMMETRIC MODEL}
\end{center}
\vspace{1cm}

%&&&&&&&&&&&&&&&&&&&&&&&&&&&&&&&&&&&&&&&&&&&&&&&&&&&&&
\begin{flushleft}{M. Maya $^1$, O. G. Miranda $^2$}\\ $^1$Facultad de Ciencias
 F\'{\i}sico Matem\'{a}ticas Universidad Aut\'{o}noma de Puebla A. P.
1364, 72000, Puebla, M\'{e}xico \\
$^2$ Centro de Investigaci\'{o}n y de Estudios Avanzados del IPN
Dpto. de F\'{\i}sica, A. P. 14-740, M\'{e}xico 07000, D. F., M\'{e}xico,
(e-mail: omr@fis.cinvestav.mx)
\end{flushleft}\vspace{1cm}

%\begin{center}{M. Maya}\\Facultad de Ciencias F\'{\i}sico Matem\'{a}ticas\\
%Universidad Aut\'{o}noma de Puebla\\A. P. 1364, 72000, Puebla, M\'{e}xico
%\end{center}               \begin{center} and\end{center}
%\begin{center}{O. G. Miranda}\\
%Centro de Investigaci\'{o}n y de Estudios Avanzados del IPN\\
%Dpto. de F\'{\i}sica, A. P. 14-740, M\'{e}xico 07000, D. F., M\'{e}xico
%\end{center}\vspace{1cm}
%&&&&&&&&&&&&&&&&&&&&&&&&&&&&&&&&&&&&&&&&&&&&&&&&&&&&&
\begin{abstract}
{We examine the decay of $Z_1$ in electrons with recent data from LEP.
The partial width $\Gamma(Z_1\rightarrow e^-e^+)$ is studied in the
framework of a left-right symmetric model with standard electroweak
corrections. Processes measured near the
resonance has served to measure the neutral coupling constants very
precisely, which is useful to set bounds on the parameters of the model.
This partial decay occurs in the resonance zone. As a consecuence
the process is independent of the mass of the additional $Z_2$
heavy gauge boson which appears in this kind of models and so we have
the mixing angle $\f$ between the left and the right bosons as the only
additional parameter. In this paper we take advantage of
this fact to set a bound for $\f$: $-9\times 10^{-3}\leq\f\leq
4\times 10^{-3}$, which is in agreement with other constraints
previously reported.}
\end{abstract}

\vspace{1cm}

\begin{center}
{\it (Submitted to {\bf Z. Phys. C})}
\end{center}

\newpage

\baselineskip=25pt
%$$$$$$$$$$$$$$$$$$$$$$$$$$$$$$$$$$$$$$$$$$$$$$$$$$$$$$$$$$$$$4
\section{Introduction}
The Standard Model (SM) [1] of the electroweak interaction between the
fermions has resisted all tests within the limits of the experimental
errors.
On the other hand there are several questions for which the SM have
not answer. One of these is the origin of the parity violation at the
present energies. The Left-Right symmetric models (LR) based on the
$SU(2)_R\times SU(2)_L\times U(1)$ group [2]
give an answer to that problem, since restore the parity
symmetry at high energies and give their violation at low energies as a
result of the breaking of gauge symmetry. At present the experiments are
finest and we have excellent measures of neutral processes so we can
put better restrictions to the parameters of the LR model.
In the present work we consider the partial decay width for the neutral
boson in an electron-positron pair, as is measured at LEP, in the framework
of the LR model with two neutral bosons: $Z_1$ which is predominantly
left and
$Z_2$ which is predominantly right and heavier. The partial width
$\Gamma(Z_1\rightarrow e^-e^+)$ can be written as a function of the
mixing angle $\f$ between $W_L^3$, $W_R^3$ and $B$ gauge bosons of the
model to give
the physical bosons $Z_1$, $Z_2$ and the photon, being $\f$ the only
extra parameter besides the SM parameters. We use the recent LEP results
[3] for the neutral coupling constant $g_{_A}$ for
constraining the mixing angle $\f$ and found that $g_{_A}$ is a good
place to look for constraints on new physics. The fact that the decay
$Z_1\rightarrow e^-e^+$ occurs at the energies near the resonance of the
$Z_1$ gives a good place for looking for new physics.\\
In the Sec. 2 we describe the model with the Higgs sector having two
doublets and one bidoublet and we find the masses of the physical bosons.
In Sec. 3 we calculate the decay rate $\Gamma(Z_1\rightarrow e^-e^+)$
including radiative corrections and using the LEP data we find the
constraint for $\f$ and in Sec. 4 we summarizes the results.

%$$$$$$$$$$$$$$$$$$$$$$$$$$$$$$$$$$$$$$$$$$$$$$$$$$$$$$$$$$$$$$$$$$$$$
\section{The LR model}
We consider a LR model having one bidoublet $\Phi$ and two doublets
$\chi_{_L}$, $\chi_{_R}$
whose vacuum expectation values break the gauge symmetry to give a mass
to the right gauge bosons heavier than the mass of the left ones.
This is the origin of the parity violation at low energies [4], that is, at
energies available at actual accelerators and reactors. The lagrangian
for the Higgs sector of the model is given by [5]
%%%%%%%%% Ec. (1)
\bq  {\cal L}_{LR}=(D_\m\chi_{_L})^\dagger(D^\m\chi_{_L})
  +(D_\m\chi_{_R})^\dagger(D^\m\chi_{_R})+Tr(D_\m\Phi)^\dagger(D^\m\Phi).
  \eq
In this lagrangian appears the covariant derivatives
%%%%%%%%% Ec. (2)
\bqn   D_\m\chi_{_L}&=&\partial_\m\chi_{_L}
      -\frac{1}{2}ig\vec{\tau}\cdot\vec{W}_L\chi_{_L}
      -\frac{1}{2}ig'B\chi_{_L},\nonumber\\
      D_\m\chi_{_R}&=&\partial_\m\chi_{_R}
      -\frac{1}{2}ig\vec{\tau}\cdot\vec{W}_R\chi_{_R}
      -\frac{1}{2}ig'B\chi_{_R},\\
      D_\m\Phi&=&\partial_\m\Phi
      -\frac{1}{2}ig(\vec{\tau}\cdot\vec{W}_L\Phi
      -\Phi\vec{\tau}\cdot\vec{W_R}).\nonumber          \eqn
Then in this model there are seven gauge bosons: $W_{L,R}^1$
and $W_{L,R}^2$ that are charged and $W_{L,R}^3$ and $B$ that are neutral.
The coupling constants for left and right sector are equal: $g_L=g_R$,
since we assume manifiest left-right symmetry [6].\\
When we introduce the vacuum expectation values of the multiplets of
Higgs, i. e.
%%%%%%%%% Ec. (3) Valores de expectacion
\bq  \left<\Phi\right>=\frac{1}{\r}\left(\begin{array}{cc}
     k&0\\0&k'\end{array}\right),\;
     \left<\chi_{_L}\right>=\frac{1}{\r}\left(\begin{array}{c}
     0\\v_L\end{array}\right),\;
     \left<\chi_{_R}\right>=\frac{1}{\r}\left(\begin{array}{c}
     0\\v_R\end{array}\right),  \eq
in the lagrangian (1), the interaction bosons get their masses.
The part of the lagrangian that contains the mass terms for the charged
bosons is
%%%%%%%%% Ec. (4) Lagrangiano de masa (cargado)
\bq  {\cal L}_{mass}^C=(\begin{array}{cc}W_L^+&W_R^+\end{array})M^C
     \left(\begin{array}{c}W_L^-\\W_R^-\end{array}\right),   \eq
where $W^{\pm}$ are the linear combinations
\[  W^{\pm}=\frac{1}{\r}(W^1\mp W^2).   \]
The mass matrix $M^C$ is
%%%%%%%%% Ec. (5) Matriz de masa (W's cargados)
\bq  M^C=\frac{g^2}{4}\left(\begin{array}{cc}
     v_L^2+k^2+k'^2&-2kk'\\-2kk'&v_R^2+k^2+k'^2\end{array}\right).  \eq
Since the process $Z\rightarrow e^-e^+$ is neutral, we fix our attention to
the mass lagrangian for the neutral sector:
%%%%%%%%% Ec. (6) Lagrangiano de masa (neutro)
\bq  {\cal L}_{mass}^N=
     \frac{1}{8}(\begin{array}{ccc}W_L^3&W_R^3&B\end{array})M^N
     \left(\begin{array}{c}W_L^3\\W_R^3\\B\end{array}\right), \eq
where the mass matrix is given by
%%%%%%%%% Ec. (7) Matriz de masa (Z's neutros)
\bq  M^N=\frac{1}{4}\left(\begin{array}{ccc}
     g^2(v_L^2+k^2+k'^2)&  -g^2(k^2+k'^2)   &   -gg'v_L^2\\
       -g^2(k^2+k'^2)   &g^2(v_R^2+k^2+k'^2)&   -gg'v_R^2\\
         -gg'v_L^2       &     -gg'v_R^2     & g'^2(v_L^2+v_R^2)
     \end{array}\right).    \eq
The mass matrices (5) and (7) are diagonalized by orthogonal
transformations. The charged mass matrix (5) is diagonalized with a
rotation which is parametrized [6] by an angle $\zeta$ which is
severely restringed [7]. The matrix that
diagonalize the neutral mass matrix $M^N$ is [8]
%%%%%%%%% Ec. (8) Matriz de mezcla neutra %%%%%%%%%
\bq  U^N=\left(\begin{array}{ccc}
     c_Wc_\f&-s_Wt_Wc_\f-r_Ws_\f/c_W&t_W(s_\f-r_Wc_\f)\\
     c_Ws_\f&-s_Wt_Ws_\f+r_Wc_\f/c_W&-t_W(c_\f+r_Ws_\f)\\
     s_W    &          s_W          &       r_W
     \end{array}\right),    \eq
with the definitions $c_W=\cos\t_W$, $s_W=\sin\t_W$, $t_W=\tan\t_W$ and
$r_W=\sqrt{\cos{2\t_W}}$, where $\t_W$ is the electroweak mixing angle.
Also $c_\f=\cos\f$ and $s_\f=\sin\f$. Here $\f$ can be considered as the
angle that mix the left and right handed
neutral gauge bosons $W_{L,R}^3$ respectively, and $B$ to give
the physical bosons $Z_1$, $Z_2$ and the photon:
%%%%%%%%% Ec. (9) Mezcla de los bosones de norma
\bq  \left(\begin{array}{c}Z_1\\Z_2\\A\end{array}\right)=U^N
     \left(\begin{array}{c}W_L^3\\W_R^3     \\B\end{array}\right).   \eq
The diagonalization of (5) and (7) gives the mass of the
charged $W^\pm_{1,2}$ and neutral $Z_{1,2}$ physical fields:
%%%%%%%%% Ec. (10) Masa del W %%%%%%%%%
\bq  M_{W_{1,2}}^2=\frac{g^2}{8}
  \left[v_L^2+v_R^2+2(k^2+k'^2)\mp\sqrt{(v_R^2-v_L^2)^2+16(kk')^2}\right], \eq
%%%%%%%%% Ec. (11) Masa del Z %%%%%%%
\bq  M_{Z_{1,2}}^2=B\mp\sqrt{B^2-4C},  \eq
respectively, with
\[  B=\frac{1}{8}[(g^2+g'^2)(v_L^2+v_R^2)+2g^2(k^2+k'^2)],   \]
\[  C=\frac{1}{64}g^2(g^2+2g'^2)[v_L^2v_R^2+(k^2+k'^2)(v_L^2+v_R^2)].   \]
Taking into account that $M_{W_2}^2\gg M_{W_1}^2$, from the expressions
for the masses of $M_{Z_1}$ and $M_{Z_2}$ we conclude that the relation
$M_{W_1}^2=M_{Z_1}^2\cos^2\theta_W$ still holds in this model.\\
To compare with experimental results [9], we introduce here a
parametrization for matrix (8) used frecuently [10, 11]. In this
parametrization the mixing angle $\theta_M$ is obtained as follows:
first, the gauge
fields $W_{3L}$, $W_{3R}$ and $B$ are transformed to interaction fields
$Z_L$, $Z_R$ and $A$. The field $Z_R$ does not couple with left-handed
currents whereas the photon $A$ interact only with the electromagnetic
current. With these conditions, the
relation between both sets of intermediate bosons is
%%%%%%%%% Ec. 12
\bq  \left(\begin{array}{c}Z_L\\Z_R\\A\end{array}\right)=U
     \left(\begin{array}{c}W_L^3\\W_R^3     \\B\end{array}\right), \eq
where
%%%%%%%%% Ec. 13
\bq  U=\left(\begin{array}{ccc}
     c_W    &      -s_Wt_W          &   -t_Wr_W\\
     0      &       r_W/c_W         &    -t_W\\
     s_W    &          s_W          &       r_W
     \end{array}\right).                          \eq

Second, the interacting fields $Z_L$, $Z_R$ and $A$ are transformed to
mass eigenstates $Z_1$, $Z_2$ and $A$. The photon do not mix
at this stage. The transformation is realized with a matrix $U'$
%%%%%%%%% Ec. 14
\bq  \left(\begin{array}{c}Z_1\\Z_2\\A\end{array}\right)=U'
     \left(\begin{array}{c}Z_L\\Z_R     \\A\end{array}\right).   \eq
The matrix $U'$ is a rotation that leave $A$ invariant:
%%%%%%%%% Ec. 15
\bq  U'=\left(\begin{array}{ccc}
     \cos\theta_M    &      \sin\theta_M          &   0   \\
     -\sin\theta_M   &      \cos\theta_M          &   0   \\
     0         &         0            &   1
     \end{array}\right), \eq
then the complete transformation is
%%%%%%%%% Ec. 16
\bq  \left(\begin{array}{c}Z_1\\Z_2\\A\end{array}\right)=U'U
     \left(\begin{array}{c}W_{3L}\\W_{3R}     \\B\end{array}\right).   \eq
We see that the matrix (8) is related to $U'U$ by
%%%%%%%%% Ec. 17
\bq  U^N=U'U,    \eq
if we set $\phi=-\theta_M$.

%$$$$$$$$$$$$$$$$$$$$$$$$$$$$$$$$$$$$$$$$$$$$$$$$$$$$$$$$$$$$$$$$$$$$
\section{The decay $Z_1\rightarrow e^- e^+$}
{}From the general lagrangian of the LR model we extract the terms for the
neutral interaction of a fermion with the gauge bosons $W^3_{L,R}$ and $B$:
%%%%%%%%% Ec. 18
\bq  {\cal L}_{int}^N=g(J^3_LW^3_L+J^3_RW^3_R)+\frac{g'}{2}J_YB. \eq
Inverting the Eq. (9) for the fields $W^3_L$, $W^3_R$ and $B$ and
inserting in (12) we find for $Z_1\rightarrow e^-e^+$ [12]:
%%%%%%%%% Ec. (19) lagrangiano efectivo
\bq  {\cal L}_{int}^N=\frac{g}{c_W}Z_1
     \left[\left(c_\f-\frac{s_W^2}{r_W}s_\f\right)J_L
     -\frac{c_W^2}{r_W}s_\f J_R\right],   \eq
where the left (right) current for the electrons are
\[   J_{L,R}=J_{L,R}^3-\sin^2\theta_W J_{em},   \]
and
\[  J_{em}=J_L^3+J_R^3+\frac{1}{2}J_Y,   \] is the electromagnetic
current. From Eq. (19) we can find the amplitude $M$ for the decay of
the $Z_1$
boson with polarization $\epsilon^\lambda$ into an electron-positron pair:
%%%%%%%%% Ec. (20) Amplitud
\bq  M=\frac{g}{c_W} \left[\bar{u}\gamma^\m
     \frac{1}{2}(\alpha g_V-\beta
     g_A\gamma_5)v\right]\epsilon_\m^\lambda, \eq
with
%%%%%%%%% Ec. (21) alfa
\bq   \alpha=c_\f-\frac{1}{r_W}s_\f,    \eq
%%%%%%%%% Ec. (22) beta
\bq   \beta=c_\f+r_Ws_\f.                \eq
If we consider radiative corrections for the standard model then we will
have
%%%%%%%%% Ec. (23) amplitud
\bq  M=\frac{g}{c_W}  \left[\bar{u}\gamma^\m
     (g_{_{V_{LR}}}-g_{_{A_{LR}}}\gamma_5)v\right]\epsilon_\m^\lambda, \eq
with
%%%%%%%%% Ec. (24) acoplamiento vectorial
\bq  g_{_{V_{LR}}}=\left[c_\f-\frac{s_W^2}{r_W}s_\f \right]
     \bar{g}_{_V}-\frac{c_W^2}{r_W}s_\f g_{_{V_R}},  \eq
%%%%%%%%% Ec. (25)  acoplamiento axial
\bq  g_{_{A_{LR}}}=\left[c_\f-\frac{s_W^2}{r_W}s_\f \right]
     \bar{g}_{_A}+\frac{c_W^2}{r_W}s_\f g_{_{A_R}},  \eq
here $\bar{g}_{_V}$ ($\bar{g}_{_A}$) is the value for $g_{_V}$
($g_{_A}$), but including radiative corrections whereas
$g_{_{V_R}}$ ($g_{_{A_R}}$) is the value for $g_{_V}$ ($g_{_A}$)
but free of radiative corrections. This is because in this kind of
models only standard model radiative corrections are taking into
account [11].\\
As we can see, in Eq. (23) we have made definitions for the vector and
axial-vector constants as effective coupling constants in the LR model.
The plots for $g_{_{V_{LR}}}$ and $g_{_{A_{LR}}}$ (Eqs. (24) and (25)
respectively)
are shown in Figures 1 and 2 as
functions of the left-right mixing angle $\f$ and the electroweak
mixing angle $s_W^2$.
We can see how $g_{_{V_{LR}}}$ has a stronger
dependence on $s_W^2$ than on $\f$, while $g_{_{A_{LR}}}$ presents the
opposite situation, a stronger dependence on $\f$ than on $s_W^2$. This
is not a surprise because in the SM at tree level $g_A$ is independent
on $s_W^2$, so the only dependence on this angle in $g_A$ is through
radiative corrections and through the LR correction, meanwhile in $g_V$
the $s_W^2$ dependence is presented already at tree level what makes it
more important. It has been noted [12] that $g_{_A}$ is a good
place to looking for deviations from the standard model at low energies.
As we can see from discussion above and from Figs. 1 and 2 this is
also valid for high energy experiments.\\
In Fig. 3 we have ploted $g_{_{A_{LR}}}$ as a function of $\f$,
where we put $s_W^2=0.2247$ which is a result which comes from
the $M_Z$ measure in the On-Shell scheme [13]. The values for $\bar{g}_{V}$
and $\bar{g}_{A}$ are given by [13]
%%%%%%%%% Ec. (26)
\bq  \bar{g}_{V}=
     \sqrt{\rho_f}\left(-\frac{1}{2}+2\kappa_f\sin^2\theta_W\right), \eq
%%%%%%%%% Ec. (27)
\bq   \bar{g}_{A}=\sqrt{\rho_f}\left(-\frac{1}{2}\right), \eq
with $\rho_f=1.0031$ and $\kappa_f=1+0.0031/\tan^2\theta_W$. The values
for $g_{V_R}$ and $g_{A_R}$ are those with $\rho_f=\kappa_f=1$.
The horizontal lines in the plot give us the experimental region of Ref.
[3]:
\[   g_{_A}^{exp}=-0.4998\pm 0.0014, \]
with a 90\% C. L..
With this experimental data from LEP for $g_{_A}$ we found for the mixing
angle between $Z_1$ and $Z_2$ the constraint
%%%%%%%%% Ec. (28)
\bq  - 9\times 10^{-3}\leq \f \leq 4\times 10^{-3},  \eq
with a 90\% C. L.. This limit is in good agreement with theoretical
results [14] previously reported. In our computation the advantage is
that the fit is independent on the mass of the $Z_2$ heavy gauge boson
because we are in the $Z_1$ resonance zone. Our result also agrees with
the experimental estimation of Ref. [9], if we take the angle $\theta_M$
of that reference as the negative of $\phi$ as was explained in Sec. 2.

%$$$$$$$$$$$$$$$$$$$$$$$$$$$$$$$$$$$$$$$$$$$$$$$$$$$$$
\section{Summary}
As a conclusion we can say that $g_{_{A}}$ is a good place for looking
for constraints on the mixing angle $\f$ and in general for new physics
because it has not a strong dependence on the electroweak mixing angle
$\sin^2\theta_W$ which has different values depending on the experiment
and on the renormalization scheme [13].
Besides, studing $g_{_A}$ in the resonance zone has the extra advantage
that the mass of an extra neutral heavy boson does not appear in the
computation, leaving only an extra parameter in the case of the LR model and
in any other model with only one additional neutral gauge boson as in
the case, for example, of the $SU(2)_L\times U(1)\times U(1)$ coming from
$E_6$ models. Further, this computation has the virtue that is necessary
only one experimental quantity, that is, the axial coupling constant
$g_A$.\vspace{0.5cm}

{\bf Acknowledgments}\newline
We would to thank Miguel Angel Soriano for
useful discussions. This work was supported in part by CONACYT-M\'{e}xico.

\newpage

\noindent{\Large{\bf Figure Captions}}
\vspace{1cm}

\noindent Fig. 1  Plot of $g_{_{V_{LR}}}$ as a function of $s_W^2$ and
$\f$. $g_{_{V_{LR}}}$ is highly dependent on $s_W^2$.
\vspace{0.5cm}

\noindent Fig. 2  Plot of $g_{_{A_{LR}}}$ as a function of $s_W^2$ and $\f$.
$g_{_{A_{LR}}}$ is almost independent on $s_W^2$ compared with the dependence
on $\f$.
\vspace{0.5cm}

\noindent Fig. 3 Plot of $g_{_{A_{LR}}}$ as a function of the LR parameter
$\f$ with the value $s_W^2=0.2247$. The horizontal lines give the experimental
 region with \\ a 90 \% C. L.

\end{document}